\documentclass[a4paper,11pt]{article}
\pdfoutput=1
\usepackage{jheppub} 
\usepackage[T1]{fontenc} 
\usepackage{amsmath,amssymb,graphicx,bbm,mathrsfs,nicefrac}
\usepackage{tikz}
\usepackage{amsmath}
\usepackage{amssymb}
\usepackage{graphicx,textcomp,float,gensymb,wrapfig, enumitem,comment,dsfont,framed,slashed,appendix,wrapfig,xcolor}
\usepackage{ bbold }
\usepackage{braket} 
\usepackage{ mathrsfs }
\usepackage[small]{caption}
\usepackage{subcaption}
\usepackage{etoolbox,hyperref,makecell}
\usepackage{feynmp}
\usetikzlibrary{arrows.meta}
\usepackage{empheq}
\DeclareMathOperator{\Tr}{Tr}

\patchcmd{\abstract}{\null\vfil}{}{}{}
\newcommand{\bea}{\begin{eqnarray}}  
\newcommand{\eea}{\end{eqnarray}}

\usepackage[utf8]{inputenc}

\begin{document}

\title{Composite Higgs models in disguise}

\author{Jack Setford}
\affiliation{Department of Physics and Astronomy, University of Sussex, Brighton BN1 9QH, UK}
\emailAdd{j.setford@sussex.ac.uk}

\abstract{We present a mechanism for \emph{disguising} one composite Higgs model as another. Allowing the global symmetry of the strong sector to be broken by large mixings with elementary fields, we show that we can disguise one coset $\mathcal G/\mathcal H$ such that at low energies the phenomenology of the model is better described with a different coset $\mathcal G'/\mathcal H'$. Extra scalar fields acquire masses comparable to the rest of the strong sector resonances and therefore are no longer considered pNGBs. Following this procedure we demonstrate that two models with promising UV-completions can be disguised as the more minimal $SO(5)/SO(4)$ coset.}

\maketitle

\section{Introduction}

The hierarchy problem is one of the most puzzling aspects of the Standard Model, and still it lacks a satisfactory solution. Composite Higgs models \cite{Kaplan:1983fs, Kaplan:1983sm, Agashe:2004rs, Azatov:2011qy} offer a fascinating explanation of the origin of the electroweak scale -- the Higgs is a composite pseudo-Nambu Goldstone boson (pNGB), which arises when a new sector becomes strongly interacting and confines. This new sector is endowed with a global symmetry, and it is the breaking of this global symmetry by non-perturbative vacuum condensates which leads to the appearance of the Higgs as a pNGB.

The low-energy behaviour of Composite Higgs (CH) models can be studied in an Effective Field Theory (EFT) framework, in which the heavy resonances of the strong sector are integrated out. This picture is useful, since we do not need to know the details of the UV-completion in order to understand the spectrum of the theory at energies below the confinement scale. The only features of the strong sector that we need to specify are its global symmetry $\mathcal G$ and the manner in which this symmetry breaks: $\mathcal G \rightarrow \mathcal H$. The pNGBs will come in a non-linear representation of the broken symmetry coset $\mathcal G/\mathcal H$, and the top-partners -- the light, fermionic resonances that are present in all realistic realisations -- will come in full representations of $\mathcal G$. A sigma-model approach then allows for a derivation of the pNGB potential (albeit in terms of unknown form-factors). In this way the main phenomenological differences between different CH models can be readily inferred from the symmetry structure of the theory.

Of course, merely plucking a symmetry out of the air is not equivalent to claiming it is \emph{realisable} in a QFT framework. Some work has been done towards constructing UV-completions of Composite Higgs models \cite{Ferretti:2013kya, Ferretti14, Ferretti:2016upr, Barnard13, Golterman:2017vdj}. Not all symmetry cosets, it turns out, were created equal. The cosets $SU(4)/Sp(4)$, $SU(5)/SO(5)$, and $SU(4)\times SU(4)/ SU(4)$ have been identified as the minimal cosets that have a UV-completion in the form of a fermion-gauge theory. The Minimal Composite Higgs Model (MCHM) $SO(5)/SO(4)$ is notably not so easy to complete. From one perspective, it might be argued that one should restrict one's attention to Composite Higgs models based on UV-completable cosets, and to take seriously the phenomenology they predict.

However in this work we describe a mechanism whereby a Composite Higgs model with the coset $\mathcal G/\mathcal H$ might, at energies currently accessible to us, be \emph{disguised} as a model with a different symmetry coset $\mathcal G'/\mathcal H'$, with $\mathcal G' \subset \mathcal G$ and $\mathcal H' \subset \mathcal H$. This can happen in such a way that at or below the confinement scale $f$, only the resonances predicted by the $\mathcal G'/\mathcal H'$ model are seen, while the remaining resonances acquire masses $\gg f$ and could remain hidden -- thus the model is disguised.

This paper is organised as follows. In Section~\ref{m}, we present a general description of the mechanism, assuming that the field responsible for deforming the strong sector is a new fermion $\psi$ which is a singlet under the SM gauge group. In Section~\ref{te}, we walk through two examples in which the original symmetry coset is $SU(4)/Sp(4)$ and $SU(5)/SO(5)$, in both cases showing that they can be disguised as the MCHM coset $SO(5)/SO(4)$. Then in Section~\ref{dwtR} we argue that the field responsible for the deformation could in fact be the right-handed top quark, if we take $t_R$ to be `mostly' composite. Finally in Section~\ref{c} we conclude our discussion.

\section{Mechanism}
\label{m}

In Composite Higgs models we assume that the new, strongly interacting sector is endowed with a global symmetry $\mathcal G$. The Higgs will be part of a set of pseudo-Nambu Goldstone bosons (pNGBs) that arise when $\mathcal G$ is spontaneously broken to a subgroup $\mathcal H$. The $n$ pNGBs live in the coset $\mathcal G/\mathcal H$, and there will be one for each broken generator, i.e. $n = \dim \mathcal G - \dim \mathcal H$. The Higgs and other pNGBs can only acquire a potential if the global symmetry $\mathcal G$ is explicitly broken by couplings to an external sector. This is normally accomplished by allowing the SM to couple to the strong sector -- these couplings then explicitly break $\mathcal G$ and induce a loop-level potential for the pNGBs.

We are going to consider a modified scenario, in which some new fields couple to the strong sector and provide an extra source of explicit breaking. We are particularly interested in the case where these new couplings are \emph{strong}. We will say that the new couplings deform, or rather, \emph{disguise} the strong sector's  symmetry properties -- due to the explicit breaking, its apparent global symmetry is now a subgroup of the original symmetry, and the pattern of spontaneous breaking has been modified.

Depending on the nature of these new fields, there are different ways they could couple to the strong sector. We are going to focus on the case where the new fields are fermionic, and couple to the strong sector via the partial compositeness mechanism \cite{Kaplan91, Contino06}. This mechanism is normally employed to allow the SM quarks (or at the very least, the top), to interact with the composite sector. Ordinarily we consider terms such as
\begin{equation}
\label{partial_compositeness}
\mathcal L \supset y_L f \overline q_L \mathcal O_L + y_R f \overline t_R \mathcal O_R + h.c.,
\end{equation}
where $q_L = (t_L, b_L)$. The $\mathcal O_{L,R}$ are composite fermionic operators with the same SM quantum numbers as $q_L, t_R$. Thus the elementary top quark mixes with the `top-partners', allowing the physical, partially composite eigenstate to couple to the Higgs.

Now, the couplings in \eqref{partial_compositeness} will explicitly break the global symmetry $\mathcal G$. If we were to write the couplings in full we would have, for instance:
\begin{equation}
\label{with_spurion}
\mathcal L \supset y_L f (\overline q_L)_\alpha (\Delta_L)^\alpha_i \mathcal O_L^i + y_R f (\overline t_R)_\alpha (\Delta_R)^\alpha_i \mathcal O_R^i,
\end{equation}
where $i$ is an index belonging to $\mathcal G$ and $\alpha$ belongs to the SM gauge group. The tensor $\Delta$ carries indices under both the SM gauge group and $\mathcal G$, parametrising precisely how the symmetry $\mathcal G$ is broken \cite{Matsedonskyi12}. One can think of $(\overline t_L)_\alpha \Delta^\alpha_i$ as an embedding of the SM top into a `spurionic' representation of $\mathcal G$. The representation into which the top is embedded should match the representation in which $\mathcal O_L$ transforms, and this ensures that the explicit breaking is treated in a way formally consistent with the symmetries of the strong sector.

As an example, let us consider the MHCM, which has the pNGB coset $SO(5)/SO(4)$. The $SO(4)$ in this model becomes the custodial $SO(4) \simeq SU(2)_L \times SU(2)_R$. We can take $\mathcal O_L$ and $\mathcal O_R$ to both be in the ${\bf 5}$ of $SO(5)$, which decomposes under the custodial group as $({\bf 2}, {\bf 2}) \oplus ({\bf 1}, {\bf 1})$. The $q_L$ then couples to the bidoublet, while the $t_R$ couples to the singlet. This translates into the following expressions \cite{Contino:2006qr} for $\Delta_{L,R}$ in \eqref{with_spurion}:
\begin{align}
\begin{split}
\Delta_L &= \frac{1}{\sqrt{2}} \begin{pmatrix}
0 & 0 & 1 & -i & 0 \\
1 & i & 0 & 0 & 0
\end{pmatrix} \\
\Delta_R &= -i\begin{pmatrix}
0 & 0 & 0 & 0 & 1
\end{pmatrix}.
\end{split}
\end{align}

Proceeding along similar lines, let us introduce a new fermion $\psi$, which mixes with a composite operator $\mathcal O_\psi$. For simplicity, let us take $\psi$ to be a singlet under the SM gauge group. The mixing terms look like:
\begin{equation}
\label{l_slashed}
\mathcal L_\slashed{\mathcal G} = y_\psi f \overline \psi \Delta_i \mathcal O_\psi^i + h.c.
\end{equation}
Note that the $\alpha$ index has been omitted, since $\psi$ is a singlet under the Standard Model.
Now we are going to assume that the mixing parameter $y_\psi$ is large -- so that $\mathcal G$ is no longer a good symmetry. Let us define $\mathcal G' \subset \mathcal G$ such that $\mathcal G'$ is the residual symmetry after the interactions with $\psi$ are included. Suppose that the global symmetry of the original theory \emph{spontaneously} breaks to $\mathcal H$, and define $\mathcal H' = \mathcal H \cap \mathcal G'$. Then, with the inclusion of $\mathcal L_\slashed{\mathcal G}$, the new theory appears to have the \emph{new} symmetry breaking pattern $\mathcal G'/\mathcal H'$. One composite Higgs model has been disguised as another.

What do we mean when we say that $y_\psi$ is large? In the language of \cite{Giudice07, Liu:2016idz}, we can broadly parametrise the strong sector via its typical mass scale $m_\rho$ and coupling $g_\rho$, which scales in large-$N$ theories \cite{Witten:1979kh} as
\begin{equation}
g_\rho = \frac{4\pi}{\sqrt{N}}.
\end{equation}
They are related to the symmetry-breaking scale via $m_\rho = g_\rho f$. The limit $g_\rho = 4\pi$ represents the limit of validity of the effective theory; for stronger couplings a loop expansion in $(g_\rho/4\pi)^2$ is no longer valid.

For $y_\psi \approx g_\rho$, the mixing angle between the elementary $\psi$ and $\mathcal O_\psi$ is large, and the physical eigenstates will have a large degree of compositeness. Operators induced by the coupling of $\psi$ to the strong sector (which violate $\mathcal G$) will be proportional to some power of $(y_\psi / g_\rho)$, and in the limit where $y_\psi \approx g_\rho$, these operators are no longer suppressed. We are justified in saying that the apparent global symmetry of the strong sector has been disguised, since operators which break the symmetry appear at the same order as operators which respect it.

In order to have a large value of $y_\psi$, we require the scaling dimension of $\mathcal O_\psi$ to be close to $5/2$. This can happen if the dynamics above the compositeness scale are approximately conformal, and the operator $\mathcal O_\psi$ has a large anomalous dimension \cite{Agashe:2004rs}. A similar requirement holds for the mixings of the top quark to the top-partners -- in order to generate a sizeable $\mathcal O(1)$ top Yukawa, the $\mathcal O_{L,R}$ must have large anomalous dimensions so that the mixing terms become effectively relevant operators.

\section{Two examples}
\label{te}

It is often remarked that the Minimal Composite Higgs model (MCHM) \cite{Agashe:2004rs} has no UV-completion in the form of a renormalisable gauge-fermion theory. As discussed in \cite{Ferretti:2013kya, Ferretti14}, a theory whose UV-completion consists of $n_i$ fermions in each representation $R_i$ of the new strongly interacting gauge group (assuming it is simple) has the following global symmetry:
\begin{equation}
\mathcal G = SU(n_1) \times \dots \times SU(n_p) \times U(1)^{p-1}, 
\end{equation}
where $p$ is the number of different irreducible representations in the model. From this we see that there is no simple gauge-fermion theory that gives rise to an $SO(5)/SO(4)$ pNGB coset.

In this section we will take two models which \emph{do} have gauge-fermion UV-completions, and show that using the above procedure they can be disguised at low energy as the $SO(5)/SO(4)$ model.

\subsection{$SU(4)/Sp(4)$}
\label{SU4Sp4}

\begin{figure}[t]
\centering
\begin{tikzpicture}
\definecolor{mycolor}{RGB}{200,200,200}
\begin{scope}
\draw[fill=white] (0,0) circle [radius=3.5] ;
\draw[fill=mycolor] (-1,0) circle [radius=2] ;
\draw[dashed] (1,0) circle [radius=2] ;
\node at (0, 3) {$SU(4)$} ;
\node at (-1.2, 1.6) {$Sp(4)$} ;
\node at (1.2, 1.6) {$Sp(4)'$} ;
\node at (0, 0) {$SO(4)$} ;
\node at (0, -2.7) {$\eta$} ;
\node at (1.8, 0) {$H$} ;
\def\cut{(8,0) circle (2.01)}
\def\big{(7,0) circle (3.2)}
\definecolor{mycolor}{RGB}{200,200,200}
\draw[fill=mycolor] (6,0) circle [radius=2] ;
\draw[white,fill=white, even odd rule] \big{\cut};
\end{scope}
\draw[dashed] (8,0) circle [radius=2] ;
\node at (8.2, 1.6) {$Sp(4)'$} ;
\node at (7, 0) {$SO(4)$} ;
\node at (8.8, 0) {$H$} ;
\draw[-{Latex[scale=2.0]}] (4,0) -- (5.5,0) ;
\end{tikzpicture}

\caption{Symmetry breaking patterns in the disguised $SU(4)/Sp(4)$ model. The solid circles represent the spontaneous breaking in the original model. The dashed circle represents the $Sp(4)'$ subgroup preserved by the explicit breaking, so that the `disguised' model becomes $Sp(4)'/SO(4)$.}
\label{breakings}
\end{figure}
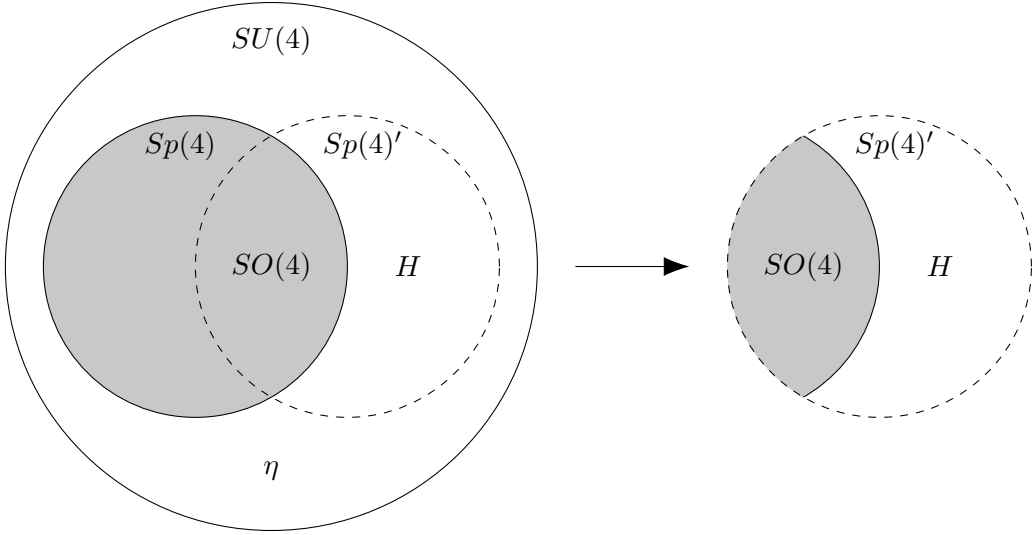

In this section we will look at the next to minimal Composite Higgs model \cite{Gripaios09, Galloway:2010bp}, in which the pNGB coset is $SU(4)/Sp(4)$\footnote{A UV-completion of this coset was studied on the lattice with an $SU(2)$ confining gauge force \cite{Hietanen:2014xca} -- the results point to a large value of $g_\rho \sim \mathcal O(10)$, in line with the large-$N$ expectation.}. This coset features one extra singlet pNGB, which we denote by $\eta$. The spontaneous breaking is achieved by a VEV in the antisymmetric ${\bf 6}$ of $SU(4)$, which we will take to be proportional to
\begin{equation}
\langle {\bf 6} \rangle \propto \begin{pmatrix}
	i\sigma^2 & 0\\
	0 & i \sigma^2\\
\end{pmatrix}.
\end{equation}
Then the pNGBs are parametrised as fluctuations around the vacuum:
\begin{equation}
\Sigma(\phi^i) = U\langle{\bf 6}\rangle U^T, \;\;\;\;\; U = \exp(i\phi^i X^i/f),
\end{equation}
where $\phi = \{H, \eta\}$ and $X^i$ are the broken generators\footnote{The calculations in this and the next section use a specific basis for the generators of $SU(4)$ and $SU(5)$. We use the bases given in \cite{Gripaios09, Ferretti14}, to which the interested reader can refer.}.

As outlined in the previous section, we will introduce a new fermionic field $\psi$, singlet under the SM. In order to disguise this model as $SO(5)/SO(4)$, we must look for a $\mathcal L_\slashed{\mathcal G}$ that explicitly breaks $\mathcal G$ to $\mathcal G' = SO(5)$. This can be done, for instance, with $\mathcal O_\psi$ in the ${\bf 6}$ of $SU(4)$. In this case \eqref{l_slashed} looks like
\begin{equation}
\mathcal L_\slashed{\mathcal G} = y_\psi f \overline\psi \Tr [\Delta \mathcal O_\psi] + h.c.
\end{equation}
The ${\bf 6}$ decomposes under $SU(2)_L \times SU(2)_R$ as:
\begin{equation}
{\bf 6} = ({\bf 2}, {\bf 2}) \oplus ({\bf 1}, {\bf 1}) \oplus ({\bf 1}, {\bf 1}).
\end{equation}
The new field $\psi$ must couple to a linear combination of the two singlets in this decomposition. The two singlets correspond to
\begin{equation}
\Delta_\pm = \begin{pmatrix}
\pm i\sigma_2 & 0 \\ 0 & i\sigma_2
\end{pmatrix},
\end{equation}
and one can verify that if we take
\begin{equation}
\label{spurion}
\Delta = \cos\theta \;\Delta_- + \sin\theta \;\Delta_+,
\end{equation}
the unbroken symmetry is indeed an $Sp(4)' \simeq SO(5)$ subgroup of the original $SU(4)$.

Notice that, using this notation, $\langle {\bf 6} \rangle \propto \Delta_+$. So long as $\theta \neq \pi/2$, the explicit and spontaneous breakings preserve \emph{different} $Sp(4)$ subgroups. That is, in our earlier notation:
\begin{align}
\begin{split}
\mathcal G' &= Sp(4)'\\
\mathcal H &= Sp(4)\\
\mathcal H' = \mathcal H \cap \mathcal G' &= Sp(4) \cap Sp(4)'.
\end{split}
\end{align}

If the spontaneous and explicit breakings preserved the \emph{same} $Sp(4)$ subgroup, then in the disguised model there would be no spontaneous symmetry breaking at all, since the spontaneously broken symmetry would never have been a good symmetry in the first place. In Fig.~\ref{breakings}, this would correspond to the $Sp(4)$ and $Sp(4)'$ circles coinciding. In such a model there would be no Goldstone bosons -- the explicit breaking leads $H$ and $\eta$ to acquire masses comparable to the other resonances of the strong sector.

Since we are trying to disguise $SU(4)/Sp(4)$ as $SO(5)/SO(4)$, we want the Higgs (but not $\eta$) to remain an exact Goldstone boson. One can verify that in the limit where $\theta \rightarrow 0$, the generators corresponding to the four degrees of freedom of the Higgs are preserved by the explicit breaking. This is the case shown in Fig.~\ref{breakings}: the Higgs lives in the part of $Sp(4)'$ which is spontaneously broken, while the $\eta$ lives in the part of $SU(4)$ which is broken by the explicit breaking, and thus acquires a large mass and is hidden. Thus we have disguised the $SU(4)/Sp(4)$ coset as $SO(5)/SO(4)$.

Note that the angle $\theta$ is parametrising some of our ignorance about the UV physics. Without having a specific UV model in mind we cannot predict the misalignment between the explicit breaking and the spontaneous breaking. With an explicit model one might be able to use lattice calculations, and/or an NJL-type analysis (see, for instance, \cite{Barnard13}), in order to obtain a better understanding of the true vacuum of the theory. For now, however, we are working at a more general level, and will treat $\theta$ as a free parameter.

Another way of seeing this mechanism at work is to look at the Coleman-Weinberg potential for the pNGBs. Including only the corrections from loops of the new fermion field $\psi$, the potential must be constructed out of invariants of $\Sigma$ and $\Delta$, i.e. it should be a function of $\Tr[\Delta^T \Sigma]$. Taking $\Delta$ as defined in \eqref{spurion}, the lowest order contribution to the CW potential is
\begin{equation}
V \propto -\Tr[\Delta^T \Sigma] \Tr[\Delta \Sigma^\dagger]
\end{equation}
\begin{equation} 
\label{pNGB_masses}
= \cos^2\theta \; \eta^2 + \sin^2\theta\;(1 - h^2 - \eta^2).
\end{equation}
We can see that in the limit $\theta\rightarrow 0$, $h$ remains an exact Goldstone boson, living in the coset $SO(5)/SO(4)$.

One should note that, in arriving at the above expression, we performed the following field redefinitions of the pNGB fields (following \cite{Gripaios09}):
\begin{align}
\begin{split}
\frac{h}{\sqrt{h^2+\eta^2}} \sin\left(\frac{\sqrt{h^2 + \eta^2}}{f}\right) \rightarrow h, \\
\frac{\eta}{\sqrt{h^2+\eta^2}} \sin\left(\frac{\sqrt{h^2 + \eta^2}}{f}\right) \rightarrow \eta.
\end{split}
\end{align}
Field redefinitions of the form $\phi \rightarrow \phi\; f(\phi)$, (with $f(0) = 1$), are valid in the context of the sigma-model \cite{Callan69}; the above redefinition is especially useful since it makes clear the fact that $h$ is an \emph{exact} pNGB in the $\theta\rightarrow 0$ limit\footnote{Furthermore, in this basis it is precisely the VEV of $h$ which sets the scale of EWSB, i.e. $m_W \propto \langle h\rangle$.}.

In order for the disguising mechanism to work, we need a small value of $\sin\theta$ -- only then will there be a hierarchy between the masses of $\eta$ and $H$. Having large values of both $\sin\theta$ and $y_\psi$ will spoil the role of the Higgs as a Goldstone boson, giving it a mass closer to that of the other strong sector resonances.

\subsection{$SU(5)/SO(5)$}

Another coset with a realistic UV-completion is $SU(5)/SO(5)$ \cite{ArkaniHamed:2002qy, Vecchi:2013bja, Ferretti14}. In this section we show that,  in complete analogy with the previous section, this model can also be disguised as the MCHM via a suitable choice of $\mathcal L_\slashed{\mathcal G}$\footnote{See \cite{vonGersdorff:2015fta} for a microscopic realisation}.

The spontaneous breaking $SU(5) \rightarrow SO(5)$ can be achieved with a VEV in the symmetric {\bf 15} of SU(5), which we take to be proportional to
\begin{equation}
\langle {\bf 15} \rangle \propto \begin{pmatrix}
	{\mathbb 1}_4 & 0\\
	0 & 1\\
\end{pmatrix}.
\end{equation}
This coset features 14 pNGBs, the Higgs, a charged $SU(2)_L$ triplet $\Phi_\pm$, a neutral triplet $\Phi_0$, and a singlet $\eta$. These are parametrised by
\begin{equation}
\Sigma = U \langle {\bf 15} \rangle U^T, \;\;\;\;\; U = \exp(i\phi^a X^a/f),
\end{equation}
but since in this case $\langle {\bf 15} \rangle$ is proportional to the identity, we can just write $\Sigma = UU^T$.

Let us assume that the new source of explicit breaking comes from a SM singlet fermion $\psi$. Then, just as before, $\mathcal L_\slashed{\mathcal G}$ is given by:
\begin{equation}
\mathcal L_\slashed{\mathcal G} = y_\psi f \overline\psi \Tr [\Delta \mathcal O_\psi] + h.c.
\end{equation}
where now we take $\mathcal O_\psi$ to be in the ${\bf 15}$ of $SU(5)$. Notice that in both this and the previous example, $\mathcal O_\psi$ was taken to be in the same representation as the operator whose VEV breaks the symmetry spontaneously.

Now the ${\bf 15}$ of $SU(5)$ decomposes under $SU(2)_L \times SU(2)_R$ as:
\begin{equation}
{\bf 15} = ({\bf 3},{\bf 3}) \oplus ({\bf 2},{\bf 2}) \oplus ({\bf 1},{\bf 1}) \oplus ({\bf 1},{\bf 1}).
\end{equation}
If we take the new source of breaking to be a SM singlet, then, just as in the $SU(4)/Sp(4)$ case, we have two singlets in the decomposition of the ${\bf 15}$ to which $\psi$ may couple. These two singlets correspond to:
\begin{equation}
\Delta_\pm = \begin{pmatrix}
	{\mathbb 1}_4 & 0\\
	0 & \pm 1\\
\end{pmatrix}.
\end{equation}
For a linear combination of the two singlets, $\Delta = \cos\theta \;\Delta_- + \sin\theta\;\Delta_+$, $SU(5)$ is explicitly broken to $SO(5)'$. Precisely as before, only in the limit $\theta \rightarrow 0$ is the Higgs untouched by the explicit breaking. Furthermore, the explicit breaking gives masses to $\Phi_\pm$, $\Phi_0$ and $\eta$. In the case where $y_\psi$ is large, the pNGB coset is disguised as $SO(5)/SO(4)$.

\section{Deforming with $t_R$}
\label{dwtR}

It has been noted \cite{Giudice07, Pomarol:2008bh, Batell:2007ez} that it is phenomenologically possible, and perhaps desirable, for the $t_R$ quark to be `mostly' composite, in the sense that $y_R$ in \eqref{partial_compositeness} is of order $g_\rho$. If this were the case, then the couplings of $t_R$ to the strong sector can indeed be thought of as changing the symmetry properties of the strong sector, and disguising the coset space as another.

Let us go back to the $SU(4)/Sp(4)$ example. Of course, unlike our hypothetical field $\psi$, $t_R$ is not a Standard Model singlet -- it is charged under $U(1)_Y$ and $SU(3)_c$. This does not change the discussion of Section~\ref{SU4Sp4}, however; we just replace $\mathcal O_\psi$ with $\mathcal O_R$, which has the same SM quantum numbers as $t_R$. In the original paper studying this coset \cite{Gripaios09}, the authors conclude that, in order to preserve the custodial symmetry that protects the $Z b \overline b$ coupling, the left and right handed quarks ought to be embedded into the $\bf 6$ of $SU(4)$ -- precisely as we did for $\psi$ in Sec.~\ref{SU4Sp4}.

It is clear that, if we want $t_R$ to couple to the Higgs and to participate in Yukawa interactions, then we must have $\theta \neq 0$. As stated earlier, we can always take $\theta$ to be small, such that a large hierarchy is generated between $\eta$ and $h$. First however, we should check that small values of $\theta$ are still consistent with a large enough top Yukawa coupling. We must embed $q_L$ into the $({\bf 2}, {\bf 2})$ of the $\bf 6$, which fixes
\begin{equation}
\Delta_L = \begin{pmatrix}
0 & Q \\
-Q^T & 0
\end{pmatrix},
\end{equation}
with $Q = (0, q_L)$. Let us assume that the couplings of $t_R$ are proportional to $\Delta_R$ in analogy to \eqref{spurion}:
\begin{equation}
\Delta_R = \cos\theta \; \Delta_- + \sin\theta \; \Delta_+.
\end{equation}
Then the Yukawa coupling of the top is obtained from the effective operator:
\begin{equation}
M_t \;\overline t_L t_R \Tr[\Delta_L^T \Sigma] \Tr[\Delta_R \Sigma^\dagger],
\end{equation}
where $M_t$ is a momentum-dependent form factor which encodes the integrated-out dynamics of the strong sector. Expanding this operator informs us that the coupling $\overline t_L t_R h$ will be proportional to $\sin\theta$.

We expect the Yukawa coupling also to be proportional to $y_L y_R$, and dimensional reasoning (discussed in detail in \cite{Matsedonskyi12}) suggests it should also be proportional to $f/m_T$, where $m_T$ is the mass of the lightest top-partner. Thus we conclude that the top Yukawa scales, up to some numeric prefactor, as
\begin{equation}
y_t \approx y_L y_R \sin\theta \frac{f}{m_T}.
\end{equation}

Furthermore, all contributions to the CW potential of the Higgs involving the right-handed top must be proportional to powers of $\Tr[\Delta_R \Sigma^\dagger]$ -- therefore the contributions to the potential must always depend on powers of $y_R \sin\theta$. In fact, the usual analyses of the size of the top Yukawa, the mass of the Higgs and the top-partners, and the required tuning for successful EWSB, proceed along all the usual lines, with the replacement $y_R \rightarrow y_R 
\sin\theta$.

The disguising mechanism relies on small values of $\sin\theta$, but of course we can make $\sin\theta$ small as long as $y_R$ is sufficiently large. The mass of $\eta$ will be proportional to $\cos 2\theta$ (from equation \eqref{pNGB_masses}), and for small $\theta$ the hierarchy between the `true' pNGB $h$ and the disguised pNGB $\eta$ is assured. Thus the couplings of the top quark alone can fulfil the requirements of the disguising mechanism.

What is the phenomenology of such a scenario? We have a set of pNGBs which couple very strongly to the top -- in this example just the $\eta$, but in the $SU(5)/SO(5)$ case we would have $\Phi_\pm, \Phi_0$ and $\eta$. In ordinary composite Higgs models we expect these extra scalars to be heavier than the Higgs by roughly a factor $\xi = v^2/f^2$. In models with around 10\% tuning, this corresponds to a mass of around $400$-$500$ GeV. In our scenario, they would be significantly heavier (how much heavier is of course dependent on the value of $\theta$, or how \emph{disguised} the model is), but their Yukawa couplings to the top would be increased by the same factor.

At sufficiently high center of mass energies, these resonances would eventually appear, along with other fermionic and vector resonances. Evidence for the disguising mechanism would be the presence of \emph{split} multiplets. For instance, in the $SU(4)/Sp(4)$ model we have top-partners in the $\bf 6$ of $SU(4)$. In the disguised model, this would be split into ${\bf 5} \oplus {\bf 1}$ of the unbroken $SO(5)$, with the singlet coupling most strongly to $t_R$. We would expect the large breaking of the $SU(4)$ symmetry to lead to a mass splitting between the five-plet and the singlet.

\section{Conclusion}
\label{c}

We have presented a mechanism whereby the symmetry breaking pattern of the strong sector can be disguised, via couplings of an elementary field to a strong sector operator. This field could be a BSM field, or, as we argued in Section~\ref{dwtR}, it could be the right-handed top quark, avoiding the need for any new fields.

This is a useful observation, especially if one has reason to believe that some pNGB cosets might be more plausible than others -- perhaps because one is concerned about UV-completions of the model. We have shown that two UV-completable cosets, $SU(4)/Sp(4)$ and $SU(5)/SO(5)$, can be deformed in such a way that at low energies the pNGB spectrum is as we would expect in an $SO(5)/SO(4)$ model.

This is certainly not equivalent to claiming that a UV-completion for the $SO(5)/SO(4)$ coset has been found. After all, the mixing $\overline\psi \mathcal O_\psi + h.c.$ will arise from a non-renormalisable operator, presumably a four-fermion operator involving $\psi$ and three techni-fermions. Nonetheless, attempts at finding a `UV-completion' of composite Higgs models so far do not speculate on the origin of these four-fermion interactions\footnote{This discussion might call into question the usage of the term `UV-completion' -- there are always problems whose solutions can be delayed to a higher scale.} (their scale can be significantly higher than the compositeness scale). Therefore it is fair to say that we have found a UV-completion of the $SO(5)/SO(4)$ coset which is \emph{just as complete} as any other composite Higgs UV-completion.

In the case where the $t_R$ is responsible for the disguise, we have a model with a set of heavy scalar resonances with very strong couplings to the top -- very strong in this case meaning close to the non-perturbative limit. We leave a detailed phenomenological analysis for future work. It would be interesting to study whether the large couplings of the scalars to the top can lead to sizable contributions to effective operators, and whether these can have any impact on Higgs or gauge boson production cross-sections.

\bibliographystyle{utphys}
\bibliography{References}

\end{document}